\documentstyle[12pt]{article}
\advance\topmargin by -1in
\advance\oddsidemargin by -0.5in

\newcommand{\npg}{\mbox{$n_{+}({\cal G})$}}
\newcommand{\be}{\begin{equation}}
\newcommand{\ee}{\end{equation}}
\newcommand{\bea}{\begin{eqnarray}}
\newcommand{\eea}{\end{eqnarray}}
\newcommand{\bean}{\begin{eqnarray*}}
\newcommand{\eean}{\end{eqnarray*}}
\newcommand{\nn}{\nonumber}
\newcommand{\alb}{\mbox{\small\boldmath$\alpha$}}

\newcommand{\raw}{\rightarrow}

\newcommand{\dvfi}{\mbox{$\partial\varphi$}}

\newcommand{\Ref}[1]{(\ref{#1})}

\newcommand{\wf}[1]{\mbox{$W^{(#1)}$}}
\def\dst{\displaystyle}

\catcode`@=11
\@addtoreset{equation}{section}

\title{\hfill {\bf {\small LANDAU-TMP-95-02}}\vskip 0.1cm
\hfill {\small hep-th/9505155}\vskip 1cm
{\bf Lattice Conformal Theories  and Their Integrable Perturbations}}
\author{{\bf Alexander V. Antonov}\thanks{Supported in part by
    International  Science
  Foundation (Grant M6N000)
and Russian Basic Research Foundation (Grant 95-02-05985a).}
\ \thanks{E-mail: antonov@landau.ac.ru} \\
{\small\it Landau Institute for Theoretical Physics,}\\
{\small\it Kosygina 2, GSP-1, 117940 Moscow V-334, Russia}\\
\and
 \fbox{\bf Alexander A. Belov}
\thanks{Supported in part by International Science Foundation (Grant
  N89000).} \\
{\small\it International Institute for Earthquake Prediction Theory}\\
 {\small\it and Mathematical Geophysics,}\\
{\small\it Warshavskoe sh., 79, k. 2, Moscow 113556, Russia}\\
\and
 {\bf Karen D. Chaltikian}
\thanks{E-mail: karen@quantum.stanford.edu}\\
{\small\it Department of Physics,}\\
{\small\it Stanford University,}\\
{\small\it Stanford, CA 94305-4060, USA}}

\def\V{{\cal V}}

\def\g{{sl}_2}

\def\half{{1\over2}}

\def\BC{{\bf C}}
\def\l{lattice }

\def\BA{{\bf A}}
\def\+{e^{\int^z\,dt\,\beta(t) \gamma(t)}}
\def\sl{\widehat{sl}_2}

\def\L{LKMA }
\def\d{\partial}
\def\<{\langle}
\def\>{\rangle}

\begin{document}
\maketitle
\begin{abstract}
We consider lattice analogues  of some conformal theories, including
WZW and Toda models. We describe discrete versions of Drinfeld-Sokolov
reduction and Sugawara construction for the WZW model.
We formulate perturbation theory in chiral sector.
We describe the Spaces of Integrals of Motion in the perturbed theories.
We  interpret the perturbed WZW model in terms
of NLS-hierarchy and  obtain an embedding of this model into the
lattice KP-hierarchy.
\end{abstract}
\thispagestyle{empty}
\vfill
\eject


\vskip 3cm
\begin{centering}
\fbox{
\begin{minipage}{15cm}
\hfill\\
\hfill\\
{\em\hspace*{0.7cm}
  This work turned out to be the last one in which Sasha Belov took
  immediate part. To our deep grief, he passed away at only 32 on
  March, 7, 1995. Up
  to his last moments he continued working and thinking about the
  problems going far beyond the scope of this paper.\\
  \indent
  \hspace*{0.7cm}
  His great  interest in developing
  this new direction and bright personality made any communication with him
  extremely interesting and enriching. Always tuned to learning  the
  latest achievements in any direction,
  he developed  his own unique style of
  thinking and approaching any problem. Unfortunately, most of his
  brilliant ideas and exciting  results  still remain in his
  notebooks. We hope most of them will be published sooner or
  later.\\
  \indent
  \hspace*{0.7cm}
  It is no doubt that continuing to work on lattice
  analogues of Conformal Theories and Integrable Models we still will be
  inspired by his ideas for a very long time.}\\
\hfill\\
\hspace*{8cm} A.~Antonov, K.~Chaltikian
\hfill\\
\end{minipage}}
\end{centering}
\vfill
\eject


\section{Introduction}
\label{intro}

In a recent series of papers by Feigin and Frenkel \cite{FF0,FF1,FF2}
a  new
approach to studying integrable systems was developed.
In particular, a new procedure of construction of quantum integrals of
motion  was proposed, based on some nice homological
constructions. It provides the universal treatment of conserved quantities
in both conformal theories and  massive integrable models, whenever
some extended symmetries are present in the theory. In many interesting
situations in quantum regime  this extended
symmetry is usually described by some Quantum Group (
see \cite{NSPQFT,FLect} for a review)
 which also appear
independently in Quantum Inverse Scattering Method \cite{qism} of study of
integrable models. The relationship between different quantum group
symmetries in one and the same model or in different but connected by
sort of limiting transition models is of great interest
nowadays. Present paper applies Feigin-Frenkel method to the lattice
analogues of some conformal theories, including WZW - model and
their  perturbations as well as associated integrable hierarchies.

There are at least three reasons for special interest in lattice
WZW model. They are mainly the same that dictate lattice reformulation of any
quantum model. But WZW  model has some peculiar features which make
these three
aspects deeply interrelated. Let us briefly sketch these reasons for
lattice setting  of WZW - model.

{\em A. Regularization.} Rigorous quantum consideration of WZW model
(as well as pure $\sigma$-model) involves renormalization for the
ultraviolet behavior of the naively formulated model is
ill-defined. The lattice regularization is the most natural one in a
sense that
it preserves the symmetries of continuous model ( or , better to say,
substitutes these symmetries by appropriate lattice analogues). Among
these symmetries
are the gauge and conformal invariance of the model. The other useful
feature of the lattice regularization is that it may be introduced
already on the classical level. Thus "quantization" results in "q-deformation"
of the model and Poisson-Lie symmetry of classical lattice appears to be
"quasiclassical limit" of quantum group symmetry. It should be mentioned
that originally \l conformal symmetry has been introduced in the framework
of analogous regularization for quantum Liouville and Toda theories
\cite{FT-86,VV,Ba,BB}.
The main contribution to formulation of the \l WZW - model has been
done by St.-Petersburg group in the papers
\cite{RST,AFST1,AFST2}. Further development of the formalism including
the first free field representation has been made by Falceto and
Gawedzki \cite{FG}.  The introduction of the \l Kac-Moody algebra (\L
) and
associated quantum lattice monodromy \cite{RST,AFST1,AFST2}
 brings us to
the next aspect of \l formulation, which is

{\it B. Unification of symmetries}, or more precisely, coupling
of the quantum group symmetries to the space-time degrees of freedom.
Probably, \L as it appears in \l WZW model is the best way of unification
of quantum-group and space-time symmetries of the model. According to ideology
of modern string theory all the symmetries of the model should be considered
on equal footing.

{\it C. Calculability}. The third essential reason for lattice setting of any
field theoretical model is calculability of the otherwise ill-defined certain
physical quantities. This is the main reason for discretization of Yang-Mills
model as well as low-dimensional quantum gravity. It is well-known that there
is a deep connection between (continuous) WZW model and topological
Chern-Simons theory \cite{Wit,MZ}.
 Certain expectation values of continuum Chern-Simons theory may be
computed as \l statistical sums defined in terms of quantum groups.

As we have seen all the three aspects of setting of quantum theory
on the \l (regularization, unification of symmetries and
calculability) are present
in WZW case. Their mutual relation originates from the fact that all of them
are intrinsically connected with quantum groups.

The aim of this work is twofold: to push forward the understanding of
lattice conformal theories, especially of \l WZW
model and to study analogues of their integrable perturbations.

The paper is organized as follows. In Sec.\ref {ffrap} on a simple
example we review the main ideas of Feigin-Frenkel approach to
the description of integrable systems.  In
Sec. \ref{formulation} we remind the St.-Petersburg definition of \l
KM algebra, introduce convenient analogue of the Chevalley basis  and
describe  the free fields representation of the \l WZW model.
 In Sec.\ref{DrS} we describe
explicitly  \l Drinfeld-Sokolov reduction and in Sec.\ref{SUG} -- \l Sugawara
construction. Then in Sec.\ref{wzw-pert} we proceed with perturbation
theory  for \l WZW model.
For the sake of simplicity we restrict ourselves with $\sl$ case.  All our
considerations are undertaken on quasiclassical
level. We describe lattice Maxwell-Bloch (MB) system. We also propose
to look  at this system as at the proper \l
analogue of NLS hierarchy.  Sec.\ref{LattKP} is devoted to study of
connection between \l
\l NLS and ``universal'' \l KP hierarchies. We find that \l NLS
hierarchy may be  understood as a special two-field realization
of the latter through a certain embedding. We also discuss \l affine
Toda theories and
describe their spaces of conservation laws. We end up with some
concluding remarks and reviewing of unsolved questions.


\section{Feigin-Frenkel approach}
\label{ffrap}

In this section we briefly review the main ideas of the cohomological approach
following the papers \cite{FF0,FF1,FF2,F1}.

Let $\pi=\oplus_i\pi_i$ be a graded phase space with Poisson structure
$\{ ,\} _{\pi}\, :\,
\pi\otimes\pi\to\pi$. The  space of the local functionals,
$${\cal F}=\oplus_i{\cal
  F}_i$$
is related to $\pi$ by the integral mapping:
 $$
\int: \pi_k\longrightarrow {\cal F}_k
$$
 Suppose there is given an action of nilpotent part $n_{+}({\cal G})$
of some semisimple  (affine) algebra $\cal G$ (resp. $\widehat{\cal
  G}$) on  $\pi_0$.
The kernel of such an action of $\npg$ as will be referred to as the
{\it Space of
  local integrals of motion} (IM).
 Traditionally, the generators
of the action of $\npg$ are called screening charges (SC).
 Thus, the space of local IM's is given by an intersection of kernels of all
SC's of the model. In finite-dimensional case it is possible to find a
set of local fields, spanning the space
\be
\label{1}
\hbox{ker}_{{\pi}_0} (\npg ).\\
\ee
Depending on the particular free field realization of SC's, one
can obtain the corresponding  $W$-algebra \cite{fatluk, fatluk1} or
Kac-Moody algebra \cite{Wak,ZWak,FFWak,BMP1}.
In the infinite-dimensional case, when SC's generate affine algebra
$\widehat{\cal G}$, we have the space  of local IM
\be
\label{2}
IM=\hbox{ker}_{{\cal F}_0}
(n_{+}(\widehat{\cal G})).\\
\ee
The most well-known examples are coming from  conformal field theory
 (\cite{dots,feld,gmsr,fatluk,Ger,BMP1}).
 and theory of non-linear integrable equations (see
\cite{frw} and references therein). In the semisimple case, with phase
space being the vertex operator algebra corresponding to some Cartan
subalgebra  we get
nothing but the Gelfand-Dickey algebra (or its quantum deformation,
$W({\cal G})$-algebra, in quantum theory). This algebra can be viewed as a
zeroth cohomology of a certain complex, resembling the
Bernstein-Gelfand-Gelfand resolution of the trivial representation of
$\cal G$. Following the same procedure in the affine case one obtains a
set   of conservation laws in involution with respect to
Gelfand-Dickey Poisson structure (or with respect to quantum
commutation relations \cite{fatluk,fatluk1,SY}).

As a simple example, consider the case $sl(2)$ in classical limit. In free
field representation the phase space
components $\pi_k$ are defined as\vspace{0.2cm}\\
$\pi_k:= \matrix{\hbox{\small Differential}\cr
\hbox{\small Polynomials}}\left (i\dvfi (x)\otimes
e^{ik\varphi }\right )$\vspace{0.2cm}\\
\noindent
where
$\varphi(x)$ is a free bosonic field with Poisson bracket
$$
\{ \varphi(x),\varphi(y)\} ={\rm sign}(x-y)
$$

 One SC
$$
Q\equiv \int dxe^{i\varphi (x)}\qquad \pi_0\stackrel{\textstyle
  Q}{\rightarrow} \pi_1
$$
makes nilpotent subalgebra of semisimple $sl(2)$. Together
with the second one

\begin{picture}(40,70)
\put(150,22){\shortstack{$\pi_0$}}
\put(160,27){\vector(3,2){30}}
\put(160,18){\vector(3,-2){30}}
\put(168,42){\shortstack{$Q$}}
\put(168,-8){\shortstack{$\overline{Q}$}}
\put(190,55){\shortstack{$\pi_1$}}
\put(190,-13){\shortstack{$\pi_{-1}$}}
\end{picture}

\bigskip
\noindent
they form nilpotent subalgebra of $\sl$ -- one actually
has to check that $Q$ and $\overline{Q}$ satisfy the
Serre's relations
$$ad_Q^3\, (\overline{Q})=0=ad_{\overline{Q}}^3\, (Q)$$
Action of the SC's is defined as the adjoint action
with respect  to the Poisson bracket ( or with respect to the
commutator in quantum
case).  It is easy to check, that
$\hbox{ker}_{\pi_0}(Q)$ is spanned by the element
$$u=\frac{1}{2}(i\dvfi (x))^2+\; i\partial^2\varphi,$$
which satisfies  the second Gel'fand-Dickey structure for the KdV
equation (or Virasoro algebra in quantum case)
$$
\{ u(x),u(y)\} =\left
(u(x)\partial_x+\partial_xu(x)-\partial_x^3\right) \delta(x-y)
$$
$\hbox{ker}_{{\cal F}_0}(Q)\cap\hbox{ker}_{{\cal F}_0}(\overline{Q})$ is
spanned by an infinite
set of commuting local functionals of $u$:
\bean
I_n=\int dx\, {\cal J}_n(u(x)),\quad n=1,2,\ldots\\
\{ I_n,I_m\} =0\\
\eean
Evolutions of $u$ with respect to the hamiltonians $I_n$ form the quantum KdV
hierarchy, while its evolution with respect to the hamiltonian
$I_0\equiv Q+\overline{Q}$ expressed in terms of field $\varphi$ gives
the sine-Gordon equation, which is a first equation in another
integrable hierarchy \cite{FTHMS}.


\section{Lattice Kac-Moody algebra and WZW model}
\label{formulation}

\subsection{Lattice Kac-Moody algebra -- St.-Petersburg definition}

In this paragraph we remind the definition of LKM due to Reshetikhin
and Semenov-Tian-Shansky \cite{RST}.
Physical application of this algebra appeared
in the papers \cite{AFST1,AFST2}, where  the idea
to consider the lattice regularization procedure was applied to the
WZW model. Authors of \cite{AFST1,AFST2} proposed the following  exchange
relations  for
the quantum lattice $L$-operator (discrete analogue of the Kac-Moody
current) was given
\bea
\label{qlkm}
J(n)_1J(n)_2=R^{+}J(n)_2J(n)_1R^{-}\nn\\
J(n+1)_1R^{-}J(n)_2=J(n)_2J(n+1)_1\nn\\
\eea
We use the standard notation $A_1\equiv A\otimes 1$, $A_2\equiv
1\otimes A$. $R^{+}$ and $R^{-}$ are the two conjugated solutions
 $$R^{-}=P(R^{+})^{-1}P$$
 of the Yang-Baxter equation (without spectral parameter)
$$R_{12}^{\pm}R_{13}^{\pm}R_{23}^{\pm}=R_{23}^{\pm}R_{13}^{\pm}R_{12}^{\pm}$$

For the $\g$ case these matrices have the following form
$$
R^+=q^\half
\left(\matrix{q^{-1}&0&0&0\cr 0&1&q^{-1}-q&0\cr 0&0&1&0\cr
0&0&0&q^{-1}}\right),\,\,
R^-=q^{-\half}
\left(\matrix{q&0&0&0\cr 0&1&0&0\cr 0&q-q^{-1}&1&0\cr 0&0&0&q}\right)
$$
For further purposes it will be more convenient for us to define
another set of variables, analogous to  Chevalley basis of $\g$.
Instead of matrix form of \L
$$
J(n)=\left(\matrix{J(n)_{11}&J(n)_{12}\cr J(n)_{21}&J(n)_{22}}\right)
$$
with "$\g$-constraint":
$$
J(n)_{11} J(n)_{22}- q^{-1} J(n)_{21}J(n)_{12}=q^\half
$$
we choose coordinates
\bea
\label{spb-chev}
e_n&=&J^{12}_nJ^{22}_n\nn\\
f_n&=&J^{21}_n(J^{22}_n)^{-1}\nn\\
h_n&=&(J^{22}_n)^2\nn\\
\eea
with exchange relations
\bea
\label{qchev}
 &&h_nh_{n+1}=qh_{n+1}h_n\nn\\
 &&h_ne_n = qe_nh_n\hspace{2.15cm}h_nf_n =q^{-1}f_nh_n\nn\\
 &&h_ne_{n+1} =qe_{n+1}h_n\hspace{1.4cm} h_nf_{n+1}=q^{-1}f_{n+1}h_n
 \nn\\
 &&e_nf_n =q^{-1}f_ne_n+q-1\hspace{0.9cm}[e_n,f_{n+1}] =(q-1)h_n\nn\\
\eea

In quasiclassical limit (with the appropriate scaling of Poisson brackets)
we obtain
\bea
\label{chev2}
\{ h_n,h_{n+1}\} &=&h_nh_{n+1}\nn\\
\{ h_n,e_n\} &=&h_ne_n \hspace{1.5cm}\{ h_n,f_n\} =(1)\nn\\
\{ h_n,e_{n+1}\} &=&h_ne_{n+1}\hspace{1.5cm}\{ h_n,f_{n+1}\} =h_nf_{n+1}\nn\\
\{ e_n,f_n\} &=& 1+e_nf_n\hspace{0.9cm}\{ e_{n+1},f_n\} =-h_n\nn\\
\eea

 \subsection{Lattice WZW model}

In some analogy with the continuous theory, one can formulate sort of
WZW model on the lattice. Attempting to  construct  the \l analogue
of the WZW lagrangian
$$
S[g]= {k\over 4\pi}\int\,tr(g^{-1}\d_+g)(g^{-1}\d_-g) dx^+\land dx_- +
 {k\over 12\pi}\int\,d^{-1}tr(g^{-1} d g)^{\land^3}
$$
one meets some principal difficulties. Instead, in papers
\cite{AFST1,AFST2,FG} it was proposed to take the classical equations
of motion
\be
\label{eomc}
\bar{\partial}(g^{-1}\partial g)=0=\partial (\bar{\partial} gg^{-1})
\ee
and fundamental Poisson bracket, discussed in \cite{AS,B,Fexch}.
$$
\{ g(x)_1,g(y)_2\} = g(x)_1g(y)_2r^{+}\theta (x-y)
+r^{-}g(x)_1g(y)_2\theta (y-x)
$$
as a starting point. Lattice analogue of eq. \Ref{eomc} is the
following difference equation
$$
g(n,\bar{n})g(n+1,\bar{n})^{-1}=g(n,\bar{n}-1)g(n+1,\bar{n}-1)^{-1}
$$
General solution of this equation has  the form
$g(n,\bar{n})=g_L(n)g_R(\bar{n})$. Considering only one chiral sector,
one can define the
corresponding \l Poisson structure for $g_L$ (or $g_R$) as
\bea
\label{gpb}
&&\{ g(n)_1,g(m)_2\} =g(n)_1g(m)_2r^{\pm}, \hbox{when
  }\left(\matrix{n>m\cr n<m}\right) \nn\\
&&\{ g(n)_1,g(n)_2\} =r^{+}g(n)_1g(n)_2+g(n)_1g(n)_2r^{-}\nn\\
\eea
Quantum version of this bracket is given by the exchange relations
\bea
\label{gq}
&&g(n)_1g(m)_2=g(m)_2g(n)_1R^{\pm},\hbox{when
  }\left(\matrix{n>m\cr n<m}\right) \nn\\
&&g(n)_1g(n)_2=R^{+}g(n)_2g(n)_1R^{-}\nn
\eea
Lattice analogue of the continuous current $J^c={k\over 2\pi} g\, \d
g^{-1}$ has the form
$$
J(n)=g(n+1)g(n)^{-1}
$$
and automaticaly obeys the exchange relations \Ref{qlkm}.

\subsection{ Lattice Wakimoto Construction}
\label{WAK}

In this paragraph we  describe the realization of
LKMA in terms of  free fields . Such a realization will be useful for
the construction of the perturbed lattice WZW model.
Recall first, that in continuous case there exists an explicit
realization of the Wakimoto module over $\hat{\cal G}$ in terms of
  $r=rank\, {\cal G}$ free scalar fields
 \bean
{\cal H}_a^{i}:\qquad [j(k)^i,j(r)^j]=k\delta^{ij}\delta_{k+r}\hspace{1cm}
[p^i,q^j]=-i\delta^{ij}\\
\eean
and $|\Delta_{+}|$ $\beta\gamma$-systems
\bean
{\cal H}_{\beta\gamma}^{\scriptstyle\alb}:\qquad [\beta (k)^{\scriptstyle\alb},
\gamma (r)^{\scriptstyle\alb}]=\delta_{k+r}\\
\eean
Such a realization was first obtained for ${\widehat{sl(2)}}_1$ by
Wakimoto \cite{Wak}, then extended to arbitrary $k$ by Zamolodchikov
\cite{ZWak} and generalized for $\widehat{sl(n)}$ by Feigin and Frenkel
  \cite{FFWak}.
Complete description of the complex of Wakimoto modules
  was given in the paper \cite{BMP1}. It was shown that the
  intertwining operators (screening charges) that build the complex
  can be  obtained from
  contour integrals of the so-called screened vertex operators.
  Authors of \cite{BMP1} showed explicitly that SC realize the
  action of the $U_q(n_{+})$ on the Fock space.
Definition of
  screening operators appears already in the theory of realizations of
  finite dimensional Lie algebras in terms of differential
  operators. Generalizations of this definition to the case of affine
  Lie algebras  are quite
  straightforward and explicit formulae for
  SC in terms of free fields can also be found
  in \cite{BMP1}.
 In the following we are going to build the lattice analogue of
 this construction.
The space of local fields
$\pi_0$  is defined as a space of finite-difference zero-degree
polynomials of  the following variables:\medskip\\
-- lattice vertex operators $a_n^i$, corresponding to the simple roots
${\alb}_i$, with exchange relations
\bea
\label{lffa}
&&a^i_na_{n+m}^j=q^{A_{ij}/2}a^j_{n+m}a^i_n,\hbox{\ for\ }m>0,\qquad
A_{ij}\hbox{ -- Cartan matrix of } sl(n)\nn\\
&&a^i_na^{i+1}_n=q^{1/2}a^{i+1}_na^i_n\nn\\
&&a^i_na^j_n=a^j_na^i_n,\hbox{\ when\ }\mid i-j\mid\geq 2
\eea

and Cartan-Weyl currents $p^i_n\equiv a_n^i(a_{n+1}^i)^{-1}$;\medskip\\
--lattice $\beta - \gamma$ systems, corresponding to the positive
roots, with exchange relations
\bea
\label{lffbg}
B_n^{\alb}\Gamma_n^{\alb}=q\Gamma_n^{\alb}B_n^{\alb}+q-1\equiv
q\xi_n^{\alb}-1\nn\\
\eea
where we denoted
$\xi_n^{\alb}=1+\Gamma_n^{\alb}B_n^{\alb}$\,\footnote{We extensively
  use the exchange relations between $\xi$ and original variables:
  $B\xi =q\xi B$, $\Gamma\xi=q^{-1}\xi\Gamma.$}.
Values of the degree function are
\bean
deg\,  a_n^i = {\alb}_i\\
deg\, B =deg\, \Gamma =0\\
\eean
The space of local functionals is defined via the
summation map
$$
\sum :\, \pi_k\raw {\cal F}_k
$$
Using the realization of the screening charges in terms of the
variables $a_n^i,\,B_n^{\alb},\,\Gamma_n^{\alb}$, one needs to
calculate the cohomology of the complex
\bean
&\pi_0 \stackrel{\textstyle \npg}{\longrightarrow}\oplus_{i=1}^l
 \pi_{{\alb}_i}\\
&deg\, \pi_{{\alb}_i}={\alb}_i\\
\eean

Below we present  explicit calculations for the
$\g$-case. Relations \Ref{lffa}, \Ref{lffbg} amount to:
\bea
\label{lffsl2}
&&a_na_{n+m}=q^{-1}a_{n+m}a_n\nn\\
&&\beta_n\gamma_n=q\gamma_n\beta_n+q-1\,\equiv\,q\xi_n-1\nn\\
\eea
Screening operators are given by the formulae:
\bean
&Q_1&\equiv {Q}_{\scriptstyle{\alb}_1}=\sum_na_n\beta_n\\
&Q_0&\equiv {Q}_{\scriptstyle {\alb}_0}=\sum_na_n^{-1}\gamma_n\\
\eean
${\alb}_1 =(1,-1)$ is the simple root of $sl(2)$, ${\alb}_0
=-{\alb}_1$ is the affine root of $\widehat{sl(2)}$.
 $Q_1$ is the single generator of $U_qn_{+}(sl(2))$, and together with
 $Q_0$ they form Chevalley basis of $U_qn_{+}(\widehat{sl(2)})$.
One finds by direct computation that the following combinations
\bea
\label{expl2}
e_n&=&\beta_n\nn\\
f_n&=&\gamma_n-q^{-1/2}\gamma_{n-1}\xi_np_{n-1}\nn\\
h_n&=&p_n\xi_n\xi_{n+1}\nn\\
\eea
obey the \L in Chevalley basis \Ref{chev2}. In continuous limit
formulae \Ref{expl2} coincide  with the  quasiclassical limit
of Wakimoto construction \cite{Wak,ZWak,FFWak}:
\bean
&&f(z)= -:\gamma\gamma\beta: (z) - \sqrt{2(k+2)} \gamma(z) H(z)- k\d\gamma(z)\\
&&h(z)= 2 :\gamma\beta: (z) + \sqrt{2(k+2)} H(z)\\
&&e(z)= \beta(z)\\
\eean
It should be mentioned that for the first time Wakimoto bosonization
on the  lattice was proposed
in the paper \cite{FG}, where the authors using the  twisted system of free
fields, constructed the lattice analogue of Bernard-Felder cohomology.
We give the general method of constructing the Wakimoto bosonization
for the LKM over the untwisted system (\ref{lffa},\ref{lffbg}) of free
fields, which seems to be simpler\footnote{There is an
  ``untwisting'' transformation, relating the $\beta\gamma$ systems
  from Ref. \cite{FG} and ours, however even after substitution we get
  different realizations of the LKM.}.


\section{Lattice Drinfeld-Sokolov Reduction}
\label{DrS}

\subsection{DS reduction in algebraic formulation ($\g$ case)}

The purpose of this paragraph is remind some essential facts about the
continuous  DS-reduction. We choose the algebraic formulation, which
can be straightforwardly put on the lattice afterwards.
In the rest of this paragraph we  follow the paper \cite{BoSc}.
Let us  introduce a pair of ghost fields $b(z),\,\,c(z)$ for the constraint
$\chi(e(z))=1$. The corresponding BRST operator has the form
$Q= Q_0+ Q_1$, where
\bean
&&Q_0= \oint {dz\over 2\pi i}\, :c\, e: (z)\\
&&Q_1= -\oint {dz\over 2\pi i}\, c(z)\chi(e(z))\\
\eean
Operators $Q_0,\,\,Q_1$ satisfy the following relations
$$
Q_0^2= Q_1^2= \{ Q_0, Q_1\}_+= 0
$$

The corresponding $W$-algebra for $\g$ case is just a  Virasoro algebra.
The energy-momentum tensor has the form
\be
\label{enmom-cont}
T(z)= T^{sug}(z)+ 2\d h(z)+ T^{gh}(z),
\ee
where $T^{sug}(z)$ is the Sugawara energy-momentum tensor
$$
T^{sug}(z)= {1\over 2(h+2)}\left(:ef+ fe:(z)+ :h^2:(z)\right)
$$
and $T^{gh}(z)= :c\d b:(z)$ is the ghost contribution.

It is well-known that the energy - momentum tensor \Ref{enmom-cont}
generates BRST-
cohomology. Following \cite{BoSc} we will show that the BRST cohomology can
be interpreted as the centralizer of screening charge. This will establish
the connection with Fateev - Lukyanov construction of $W$-algebras in
terms of free bosons \cite{fatluk1,fatluk}. It should be mentioned that all
the consideration
can be naturally generalized for arbitrary W${\cal G}$-algebra.
The realization of the BRST cohomology in terms of Fateev - Lukyanov
approach is extremely useful for our further \l consideration.
In fact, B. Feigin's approach \cite{F92} to construction of \l $W$-algebras
is based on direct lattice setting of Fateev- Lukyanov construction.
Lattice $W$-algebra is defined as a set of local lattice fields
generating the intersection of the kernels of appropriate system of
screening charges (Feigin - Fuks operators).

The idea of reinterpreting of BRST-cohomologies in terms of screening
charges construction is based on  using  the  spectral sequence
technique \cite{BottTu}. Let us apply this technique to the double complex
with horizontal differential $Q_0$ and vertical differential $Q_1$.
In general, spectral sequence is an instrument of calculation
$Q= Q_0+ Q_1$-cohomologies in the framework of iterative procedure
starting with $Q_0$-cohomologies and improving them "perturbatively"
on each step of certain procedure of construction a successive series
of differential complexes converging to $Q$-cohomologies. In our
case , however, spectral sequence collapses after the second
correction, which gives the result
$$
H_Q(*)\simeq H_{Q_1}(H_{Q_0}(*)).
$$
It can be proved that
$Q_0$-cohomology is generated by the fields $\tilde{h}(z)$ and
$c(z)$, where
$$
\tilde h(z)= \alpha_+ h(z)+ 2 :b\, c:(z)
$$
However, the algebra of $c$ and $\tilde{h}$ in cohomologies is not free.
The field $N(z)= \d c(z)- :c\,\tilde h:(z)$ turns out to be
$Q_0$-exact.
After
factorization over this  "null- field" $N(z)$ one can identify in
$Q_0$-cohomology
\bean
&&c\,\longrightarrow\,\V= e^{- i \alpha_+\phi(z)}\\
&&\tilde{h}\,\longrightarrow\,H(z)= i \d\phi(z)\\
\eean
Of course, $c\,c$  OPE sector differs from $\V\,
\V$ one, but this  is inessential for the calculation of
$ H_{Q_1}(H_{Q_0}(*))$.
As there are no $Q_0$-cohomologies at negative ghost numbers, one has
$$
H^{(0)}_{Q_1}= {\rm Ker} Q_1={\rm Ker}\oint {dz\over 2\pi i} \V(z)
$$
on the space of normally ordered differential polynomials of $H(z)$.
The result is given by
$$
T(z)= \half : H^2: (z)- (\alpha_+ + \alpha_- ) \d H(z).
$$
This $W_2$-algebra ($=$ Virasoro in our case) comes  naturally equipped
with realization in terms of free bosonic field (without any relation
of  Wakimoto or some other bosonization of Kac-Moody currents).

\subsection{Lattice Drinfeld-Sokolov Reduction}
\label{DS}

In this paragraph we generalize the previous procedure for the \l
case.
We introduce \l ghost system $b_n,\,\,c_n$ with relations
$$
\{b_n,\,\,c_m\}= \delta_{n m}
$$
and \l BRST-operator $Q= Q_0+ Q_1$, with
$$
Q_0= \sum_n c_n e_n ,\quad Q_1= \sum_n c_n.
$$
Applying spectral sequence technique to the double complex, built on
$Q_1$ and $Q_0$ we obtain quite analogously to continuum case that
spectral sequence terminates on the second step
$$
H_Q(*)\simeq H_{Q_1}(H_{Q_0}(*)).
$$

The $Q_0$- cohomologies are generated by \l fields
$\tilde h_n$ and
$c_n$, where
$$
\tilde h_n= h_n ( 1- b_n c_n )( 1- b_{n+1} c_{n+1} ).
$$
As its continuum counterpart the Poisson algebra for the fields
$\tilde h_n$ and
$c_n$ is not free in $Q_0$-cohomologies for there exists \l field
$N_n= c_n \tilde h_n- c_{n+1}$, which is $Q_0$-exact. The resulting Poisson
algebra (factorized by the "null-vector" $N_n$) is isomorphic to the Poisson
algebra  for \l "vertex operator" $a_n$ and \l $U(1)$- current $p_n$
(namely, it reproduces the latter algebra in $a-p$ and $p-p$ sectors
after identification $c_n \leftrightarrow a_n,\,\,
\tilde h_n \leftrightarrow p_n$ ). Thus calculation of $Q_1$- cohomologies
of complex $Q_0(*)$ reduces to Feigin's construction of the lattice $W$-algebra
(\l Virasoro or Faddeev-Takhtadjan-Volkov (FTV) algebra in our $\sl$- case ).
Namely, we should find the kernel of screening operator
$$
Q_1= \sum_n a_n
$$
This coincides with the cohomological definition of the \l Virasoro
generator \cite{F92}.

Now we can calculate $Q_1$-cohomological class in  $Q_0$-cohomologies, i.e.
we should find an element constructed from $c_n,\,\tilde h_m$ such that
it is closed (and not exact) up to some null-field (\l field containing
$N_n= c_n \tilde h_n- c_{n+1}$ at least once). The answer is given in
\cite{F92}. The following \l field
$$
\tilde A_n= {1\over (1+\tilde h_n) (1+\tilde h_{n+1}^{-1})}
$$
is an appropriate cohomology class
$$
\{ Q_1, \tilde A_n \}= -\hat N_n- \hat N_{n+1},
$$
where $\hat N_n$ is a null- field of the form
$$
\hat N_n= {N_n\over 1+\tilde h_n}.
$$
It is easy to check  that $\tilde{A}_n$ forms FTV algebra
\bea
\label{ftv}
&&\{ \tilde{A}_n, \tilde{A}_{n+2} \}= \tilde{A}_n \tilde{A}_{n+1}
\tilde{A}_{n+2}\nn\\
&&\{ \tilde{A}_n, \tilde{A}_{n+1} \} = \tilde{A}_n \tilde{A}_{n+1}
( - 1+ \tilde{A}_n+ \tilde{A}_{n+1})\nn\\
\eea
It is also easy to verify that null-fields form an ideal in the
Poisson algebra of
fields constructed from $c_n$ and $\tilde{h}_n$. This is a consequence of the
following relations
\bean
&&\{ c_{n+1}, N_n \}= - c_n N_n\\
&&\{ \tilde h_n, N_n \}= \tilde h_n N_n\\
&&\{ \tilde h_{n-1}, N_n \}= \tilde h_{n-1} N_n\\
&&\{ N_n, N_n \}= -(c_{n-1}+ c_n ) N_n\\
\eean
Notice that $N_n$ is a "fermi-field", i.e. $N_n^2= 0$.

One may be interested in construction of cohomological class $B_n$ which
is ``better'' than the found one $\tilde{A}_n$ in such a way that
$\{ Q, B_n\} =0$. In other words, $B_n$ should represent the
cohomology class of a double
complex on the original phase space, not factorized over the
null-field. In next section we will consider the Sugawara construction
as an example of such a class. Here we explain how to organize the
``improvement'' process.
The idea of construction of the class $B_n$ is to find such corrections to
$ln A_n$ which kill $\hat N_n$ terms. This can be done with the help
of a staircase sequence in the double complex.  Consider the following
sequence
\vskip 0.5cm

\noindent
\begin{picture}(340,180)(0,0)
\put(3,180){0}
\put(0,120){$\ln \tilde{A}_n$}
\put(4,130){\vector(0,1){45}}
\put(30,125){\vector(1,0){45}}
\put(80,120){$-(\tilde{N}_n+\tilde{N}_{n+1})$}
\put(80,60){$-(\phi_n+\phi_{n+1})$}
\put(115,75){\vector(0,1){42}}
\put(155,65){\vector(1,0){50}}
\put(215,60){$\phi_n\tilde{N}_n+\phi_{n+1}\tilde{N}_{n+1}$}
\put(230,0){$\dst\frac{1}{2}\phi_n^2+\frac{1}{2}\phi_{n+1}^2$}
\put(260,15){\vector(0,1){40}}
\put(305,7){\vector(1,0){40}}
\put(350,0){$\ldots$}
\put(6,150){\small$\{Q_0,\bullet\}$}
\put(35,112){\small$\{Q_1,\bullet\}$}
\put(118,90){\small$\{Q_0,\bullet\}$}
\put(167,52){\small$\{Q_1,\bullet\}$}
\end{picture}
\vskip 0.5cm
\noindent
where $\dst
\phi_n=\frac{\tilde{f}_{n+1}}{1+\tilde{h}_n}$.
After the summation of this (infinite) staircase process one obtains
\bean
&&ln B_n= ln A_n+ \sum_{m=1}^\infty {\phi_n^m+ \phi_{n+1}^m\over m}
\,\,\,\hbox{ or }\\
&&B_n = {A_n\over (1-\phi_n)(1-\phi_{n+1})}\\
\eean
Through direct computation one verifies that $B_n$ commutes with
$Q=Q_0+Q_1$ and
obeys the same FTV - algebra  \Ref{ftv} as
$\tilde{A}_n$ does.


\section{Lattice Sugawara Construction}
\label{SUG}
In this section we are going to discuss the analogue of the Sugawara
construction on the lattice. The question of what object is to be
considered an analogue of the Sugawara element is rather ambigious,
because it is not exactly clear what invariant property tells us that
some element is the Sugawara-like one. Before we proceed with
calculations, we make one comment, concerning the classical case. In
continuum, the Sugawara element satisfies the second Gel'fand-Dickey
Poisson algebra with {\em zero} central charge\footnote{In quantum
  case, however, the
central charge becomes non-zero due to quantum corrections.}. On the
other hand the
continuous limit of FTV algebra \Ref{ftv} reproduces the classical
Virasoro algebra but with {\em
  non-zero} central term. This makes one believe that the generator
of the FTV algebra should contatin some twisting part in the
continuous limit independent on the underlying algebra it is built of.
In the course of DS reduction such a twisted energy - momentum appears
naturally and is given by
\be
\label{tensor-sug}
T(z)= {1\over 2(h+2)}: J^+ J^- + J^-J^+ +\half J^0 J^0 : + \half \d J^0+
:\d b\,c: .
\ee
Below  we construct such a class $A^{\rm sug}_n$ which
coincides in the  continuous limit with \Ref{tensor-sug}.

For this purpose we start with a
$$
A_n= {1\over (1+h_n)(1+h_{n+1}^{-1})}
$$
and after summation of a certain staircase  process (slightly more complicated
than the one constructed above) obtain the desired class
\be
\label{ftv-sug}
A^{\rm sug}_n= {1\over (1+h_n+\xi_n)(1+h_{n+1}^{-1}+\eta_{n+1})},
\ee
where $\xi_n$ and $\eta_n$ are net corrections (after summation of a
staircase process). The explicit form of $\xi_n$ and $\eta_n$ is
\bean
&&\xi_{2n}= x_{2n}, \qquad \xi_{2n+1}= h_{2n+1} y_{2n+1}\\
&&\eta_{2n}= y_{2n}, \qquad \eta_{2n+1}= h_{2n+1}^{-1} x_{2n+1}\\
\eean
and
$$
x_n= e_n f_{n+1}+ (b_{n+1}- b_n) c_{n+1},\qquad
y_n= e_{n+1}f_{n+1} h_n^{-1}+ (b_{n+1}- b_n) c_n.
$$

It is easy to see that the field \Ref{ftv-sug} obeys FTV - algebra
\Ref{ftv}  and in the
continuous limit (in the leading nontrivial order of a \l spacing
$\Delta$) reduces to the classical limit of \Ref{tensor-sug}.
After suppressing ghost fields ($b_n=c_n=0$) one obtains twisted  \l
Sugawara  element.  Naturally, $A_n^{\rm sug}[b_n=c_n=0]$ obeys the same
FTV algebra.

In the end of this section, let us remind, that there exists another
realization of the FTV algebra in terms of \L \cite{BC3}. It has in
somewhat the similar form
$$
A_{2n}=\frac{1}{M_n^0}\qquad\quad A_{2n+1}=\frac{1}{M_n^1}
$$
where
\be
\label{M-fields}
M_n^p=\frac{e_nf_{n+p}}{h_nh_{n+1}\ldots h_{n+p-1}}
\ee
This realization turns out to be interesting in connection with lattice KP
hierarchy, discussed in Section \ref{lkp}. Here we only would like to
notice, that the fields $M_n^p$ obey the nice {\em Lie} algebra of the
$w_{\infty}$ - type.


\section{Perturbed \l WZW model}
\label{wzw-pert}
\subsection{Formulation of the Model}

In this section we describe the construction, to which we refer to as
 "\l perturbed WZW model" having in mind
the  parallelism with
continuous case \cite{ABF}. As in the Section \ref{formulation} we
will not construct any Lagrangian
perturbation theory (in $1+1$ \l space), but rather  consider
Hamiltonian  perturbation in "one chiral sector" of the \l WZW model.

We begin with construction of a lattice analogue of $\oint \Phi_{1 0}(z)\,{d
  z\over 2\pi i}$ perturbation operator where $\Phi_{j m}$ is a
primary field of spin $j$ with projection $m$. We choose
$$
Q_0= \sum_n a_n^{-1}\gamma_n .
$$
for that purpose.
To form nilpotent subalgebra
$n_+=e\oplus\g\otimes\BC[[t]]$ of affine algebra $\hat{sl_2}=\g\otimes\BC(t)$
we add another screening
$$
Q_1= \sum_n a_n\beta_n
$$
We introduce "$deg$" operator
$$deg\, a_n=1\qquad deg\, a_n^{-1}=-1\qquad deg\,\beta_n=deg\,\gamma_n=0$$
 so that
$deg\,\, Q_0= -1$,
$deg\,\,Q_1= 1$ and the ``correct'' adjoint action as improved Poisson brackets
$$
ad_A B:= \{A, B\}- (deg A \,\,deg B) A B.
$$
Then we have  Serre relations
$$
ad_{Q_0}^3 Q_1= ad_{Q_1}^3 Q_0= 0
$$
fulfilled, so that the definition of perturbation operator is correct.

Let us  consider now dynamical system with phase space of \l fields of
zero degree
constructed from \l
$\beta-\gamma-a$ - system. Hamiltonian of the system is
\be
\label{hamilt}
H= Q_0+ Q_1.
\ee
The system corresponds to perturbed "chiral sector" of the \l WZW model.

Our purpose now is to prove the integrability of this system and calculate
the integrals of motion (IM) following the analogy with the continuous case,
where the  system with Hamiltonian \Ref{hamilt} coincides with the
Maxwell-Bloch eq. \cite{ABF}.
We will also give interpretation of the model in terms of \l analogue of NLS
hierarchy.

Let us start from an observation that the field $h_n$
is a "zero mode" because it is conserved under the system evolution:
$$
ad_{H}(h_n)=0
$$
This implies  it is necessary to reduce our dynamical system and exclude
the field $h_n$.  We introduce new \l fields
$$
x_n= \beta_n a_n,\quad y_n= \gamma_n a_n^{-1}.
$$
Corresponding Dirac brackets for these fields (up to a
sign change) are
\bea
\label{xyPB}
&& \{x_n, x_m\}_D= -{\rm sign}(n-m) x_n x_m,\nn\\
&& \{y_n, y_m\}_D= -{\rm sign}(n-m) y_n y_m,\nn\\
&& \{x_n, y_m\}_D= {\rm sign}(n-m) x_n y_m- \delta_{n m}(1+ x_n y_m)\nn\\
\eea
In these variables Hamiltonian has the form
$$
H= \sum_n{(x_n+ y_n)}.
$$
 Poisson algebra \Ref{xyPB} strongly reminds that of from the
Feigin- Enriquez model (FE) \cite{FE}. The only difference is the "central
extension" ($\delta_{n m}$ term ) in $x-y$ sector. This central term
changes IM and dynamics drastically.
Nevertheless, cohomological structure of the space of IM appears to be
rigid with respect to
such a deformation of Poisson brackets.

\subsection{Interpretation of the model in terms of the NLS hierarchy}

Before systematical study of IM we give a brief description of our model in
terms of \l analog of Nonlinear Schrodinger (NLS)- hierarchy.
Renaming the variables
$ e_n\,\equiv\,\psi_n,\quad f_n\,\equiv\,\bar\psi_n$ for better similarity
we find that eqs. \Ref{xyPB} reproduce exactly the first Poisson
structure of the \l analogue of NLS hierarchy.
\bean
&&\{\psi_n, \psi_m\}= -{\rm sign} (n-m) \psi_n \psi_m\\
&&\{\bar\psi_n, \bar\psi_m\}= -{\rm sign}(n-m) \bar\psi_n \bar\psi_m\\
&&\{\psi_n, \bar\psi_m\}= -{\rm sign}(n-m) \psi_n \bar\psi_m
-(1+ \psi_n \bar\psi_n) \delta_{n,\,m}+ \delta_{m,\,n+1 }\\
\eean

The pair of brackets \Ref{xyPB} and
$\{\,\,,\,\,\}_0$ defined by
$$
\{ e_n, f_{n+1}\}_0= 1
$$
together with the two integrals
\bean
&&I_0= \sum_n\,ln\,(1+ \psi_n \bar\psi_{n})\\
&&I_1= \sum_n\,(\psi_n \bar\psi_{n+ 1})\\
\eean
define a bihamiltonian system.
Having  bihamiltonian structure one can prove the existence of an
infinite family of IM in involution. In next section  we
will give another proof of this fact based
on consideration of deformed FE model.
In continuum limit correspondence between few first IM's  and their flows
is as follows
\bean
&&I_0\rightarrow N \hbox { (particle number)},\\
&&I_1- I_0\rightarrow P \hbox{ (momentum)},\\
&&I_2- 2 I_1+ I_0 \rightarrow \hbox { (NLS Hamiltonian)},\\
\eean
where
$$
I_2= \sum_n \left({\psi_n^2 \bar\psi_{n+ 1}^2\over 2}-
\psi_n \bar\psi_{n+ 2}\right)
$$

\subsection{Hidden FTV algebra}

There is an interesting hidden symmetry on the reduced phase space.
Namely, the variables $ N_n^0= e_n f_n= \psi_n \bar\psi_n$
and $N_n^1=e_n f_{n+1}= \psi_n \bar\psi_{n+ 1}$ obey FTV
algebra \Ref{ftv} under the following identification
\be
\label{strange-ftv}
A_{2n}= {1\over N_n^0},\quad A_{2n+ 1}= {1\over N_n^1}.
\ee
Let us mention about strange ``notational'' similarity here. If $e_n$ and
$f_n$ were the fields from the Chevalley basis of \l $\g$, one would
have the same formulae \Ref{strange-ftv} for the generators of the FTV
algebra on the phase space reduced under $h_n=1$ constraint
\cite{BC3}.

\subsection{Deformed FE model}

In this section we will give an alternative proof of the existence
of an infinite system of IM for our model with Hamiltonian \Ref{hamilt}.
Consider the following one-parameter deformation of FE model
\bea
\label{fe-deform}
&& \{ x_n, x_m\}_D=-{\rm sign}(n-m) x_n x_m\nn\\
&& \{ y_n, y_m\}_D= -{\rm sign}(n-m) y_n y_m\nn\\
&& \{ x_n, y_m\}_D= {\rm sign}(n-m) x_n y_m- \delta_{n m}(\lambda+ x_n
y_m)\nn\\
\eea
with Hamiltonian
$$
H= Q_+ + Q_-,
$$
where $Q_+= \sum_n x_n$ and $Q_-= \sum_n y_n$. For the \l variables
$x_n$ and $y_n$ we have
$$
deg\,x_n= 1,\qquad deg\,\,y_n= -1\,.
$$
For $\lambda= 0$ we obtain FE model and for $\lambda= 1$ we come to
our initial algebra \Ref{xyPB} corresponding to perturbed \l WZW ( or
\l NLS) model. It should be mentioned that all the Poisson algebras
${\cal A_\lambda}$ defined by bracket \Ref{fe-deform} are pairwise isomorphic
for $\lambda\in (0,\,\infty)$.

In the paper \cite{FE} IM for the system \Ref{fe-deform} for $\lambda=
0$ have been
expressed in terms of cohomology classes. Standard arguments
give that the ring of cohomologies does not change under the infinitesimal
variation of basic algebraic structure \Ref{fe-deform} on the phase space.
The isomorphism of algebras ${\cal A}_{\lambda\neq 0}$ allows us to
replace an infinitesimal deformation by the finite one. Thus, the rings of
cohomologies for FE model and perturbed \l WZW model are the same.


\section{Embedding of the lattice NLS hierarchy into the lattice KP hierarchy}
\label{LattKP}
\subsection{Nonlinear Lattice $W_{\infty}$ algebra .}
\label{lwinf}

We briefly remind how Feigin's construction of $LW_N$ - algebra can be
extended to the case of $N=\infty$ \cite{BC2}. We restrict ourselves
with quasiclassical case.
Let us introduce the following set of \l variables $\{ a_n^j\}_{j=1}^N$
with the following Posson structure
\bea
\label{aabrak}
&&\{ a_n^i,a_m^i\} = {\rm sign}(m-n) a_n^ia_m^i\nn\\
&&\{ a_n^i,a_n^{i+1}\} =-\frac{1}{2}a_n^ia_n^{i+1}\nn\\
&&\{ a_n^i,a_m^{i+1}\} =-\frac{1}{2}{\rm sign}(m-n) a_n^ia_m^{i+1}\nn\\
\eea
Values of the degree function on these fields are as usual
$$\dst deg(a_n^i)=1\,,\quad deg((a_n^i)^{-1})=-1\,\qquad i=1,..., N-1
$$
Denote by $\Pi_n$ the space of finite difference polynomials of degree
$n$.
Following the papers \cite{F92,BC1,BC2,PK} we
define lattice $W_N$ algebra ($LW_N$) as an intersection
\be
\label{kernels}
\dst\cap_{i=1}^{N-1}{Ker(ad_{Q_i})}\cap \Pi_0
\ee
where $\dst Q_i=\sum_n{a_n^i}$ are the corresponding {\em screening
  charges}.
This intersection is known to be spanned by $N-1$ generators
$L_n\equiv W_n^{(2)},\,W_n^{(3)},\ldots ,W_n^{(N)}$. In the limit $N\to\infty$
we  obtain
lattice analogue of classical non-linear $W_{\infty}$-algebra. Now let
us complete this program explicitly. To construct the generators
$W_n^{(i)}$ we will use more convenient variables than $a_n^i$. First
of all, we exclude no-zero degree components of the phase space by
using as the basic variables the lattice analogues of the Cartan
currents of $sl_N$, associated with simple roots ${\bf \alpha}_1$,
${\bf\alpha}_2$, $\ldots$:
$$\dst
p_n^i=a_n^i{(a_{n+1}^i)}^{-1}
$$
Calculations with these variables turned out to be rather tedious
\cite{BC1}, so this time we choose another basis
in the root system of $sl_N$ and use Weyl chamber generators ${\bf
  \alpha}_1$, ${\bf \alpha}_1+{\bf \alpha}_2$, ${\bf \alpha_1}+{\bf
  \alpha}_2+{\bf \alpha}_3$, $\ldots$ :
\bea
\label{weylcurr}
&&k_n^1=p_n^1\nn\\
&&k_n^2=p_n^1p_n^2\nn\\
&&k_n^3=p_n^1p_n^2p_n^3\nn\\
&& \ldots\,\dots\,\ldots,\nn\\
&&k_n^{N-1}=p_n^1p_n^2\ldots p_n^{N-1}\nn\\
\eea
The following combinations turn out to be the best for our purposes:
\bea
\label{alphafields}
&&\alpha^1_n=\sum_{i=1}^{N-1}k^i_n\nonumber\\
&&\alpha^2_n=\sum_{i=1}^{N-1}\sum_{j=i+1}^{N}k^i_{n}k^j_{n+1}\nonumber\\
&&\ldots\,\ldots\,\ldots\nn\\
&&\alpha^p_n=\sum_{\{ u(l)\} }\prod_{j=1}^{p-1}k^{u(j)}_{n+j}\nonumber
\eea
where the summation goes over all sets $\{ u(j)\}$ such that
$u(j+1)>u(j)$. They form a quadratic Poisson algebra
\be
\label{w-inf-1}
\{ \alpha_n^{p}, \alpha_{n+m}^q\}_1= \theta_m^{p\,q}(- \alpha_n^{p}
\alpha_{n+m}^q +
\alpha_n^{q+m} \alpha_{n+m}^{p-q}),
\ee
where $\dst\theta_m^{p\,q}=\theta (p-m)\theta(q-p+m-1)$ and $\theta(x)$ is a
 step function $\dst\theta(x)= \left\{ \begin{array}{ll} 1, & \,\,x\geq
0\\ 0, & \,\, x<0\end{array}\right.$.
The index 1 of the bracket indicates that this is the analogue of
{\em first} Gelfand-Dickey (GD) structure for the $N$-KdV
hierarchy. Generators of $LW_N$ form the analogue of the {\em second}
GD structure. It turns out that quite parallel to the continuous case
there are Miura transformation, relating the $\alpha$ - fields
\Ref{alphafields} with the generators of $LW_N$. Direct calculation
(the most convincing method of proof) shows that the following fields
commute with all screening operators ($i=2,3,\ldots,N$):
\begin{eqnarray}
\label{mNN}
&&\wf{i}_n=\frac{\alpha^{i-1}_{n+1}+\alpha^i_n}{(1+\alpha^1_n)\ldots
(1+\alpha^1_{n+i-1})},\qquad i=2,3,\ldots, N-1 \nonumber\\
&&\wf{N}=\frac{\alpha^{N-1}_{n+1}}
{(1+\alpha^1_n)\ldots (1+\alpha^1_{n+N-1})}\hspace{5cm}
\end{eqnarray}
In the limit $N\to \infty$ one find the brackets between the fields
$\wf{i}$. Putting $\wf{1}\,\equiv\,1$, we have


\bea
\label{w-infinity}
&&\{ W_n^{p}, W_{n+m}^q\}= W_n^{p} W_{n+m}^q ( 1- W_{n+m-1}^q-
W_{n+p}^q)-W_n^{q+m} W_{n+m}^{p-m}\nn\\
&&\hspace*{2.5cm}  - W_n^{p+1} W_{n+m}^{q}+
W_n^{p} W_{n+m-1}^{q+1},\hspace*{1cm} \hbox{for } m\leq p,\,\,q+m\geq
p+1,\nn\\
&&\{ W_n^{p}, W_{n+m}^q\}= W_n^{p} W_{n+m}^q ( - W_{n+m-1}^q+
W_{n+m+p}^q)\nn\\
&&\hspace*{2.5cm} -W_n^{p} W_{n+m-1}^{q+1} - W_n^{p} W_{n+m}^{q+1},
\hspace*{1.3cm} \hbox{for } m\geq 1,\,\,p\geq m+q+1.\nn\\
&&\{ W_n^{p}, W_{n+p+1}^q\}= -W_n^{p} W_{n+p}^1  W_{n+p+1}^q- W_n^{p+q+1}+
 W_n^{p+1} W_{n+p+1}^{q}+ W_n^{p} W_{n+p}^{q+1},\nn\\
&&\{ W_n^{p}, W_{n+p-q}^q\}= -W_n^{p} W_{n+p-q-1}^1  W_{n+p-q}^q
 W_n^{p} W_{n+p-q-1}^{q+1},\qquad \hbox{for } p\geq q+1,\nn\\
&&\{ W_n^{p}, W_{n}^q\}= -W_n^{p} W_{n+p}^1  W_{n}^q
 W_n^{p+1} W_{n}^{q},\hspace*{3.6cm} \hbox{for } q\geq p+1,\nn\\
\eea
Notice that  algebra \Ref{w-infinity} is {\em unhomogeneous}
with respect to natural
gradation:  if we ascribe to the \l field $deg W_n^p= p$, we find
that the r.h.s. of \Ref{w-infinity} has degree greater or equal  than l.h.s.
It is possible, however, to split the bracket \Ref{w-infinity} in two
parts  $\{\,,\,\}=
-\{\,,\,\}_1+\{\,,\,\}_2$, where $\{\,,\,\}_1$ is defined by
\Ref{w-inf-1} and $deg \,\{\,,\,\}_1=0,\,\,deg\,\{\,,\,\}_2=1$.
Ending this section, we would like to highlight several points:
\begin{description}
\item - Distinct from continuous case, for any {\em finite} $N$, algebra
  $LW_N$  does not form a
  subalgebra of $LW_{\infty}$. However, by forcing $\wf{i}=0$ for
  $i\geq N$ one can obtain any $LW_N$.
\item - In continuous case there exists the so-called {em two-boson}
  realization of KP hierarchy \cite{yuwu}, in which
  $W_{\infty}$-algebra generators are expressed in terms of two $u(1)$
  currents. Analogous construction happens to exist on the
  lattice. Fields forming Poisson algebra \Ref{w-infinity} can be
  realized in terms of two lattice $u(1)$ currents \cite{BC2} $u_n=t_{2n}$
  and $v_n=t_{2n+1}$, commuting as
\be
\label{ftv-1}
\{t_n,t_{n+1}\}\,=\,-t_nt_{n+1}
\ee
\item - Under properly defined continuous limit the brackets 1 and 2 become
  the corresponding Poisson structures of the KP-hierarchy
  (resp. linear $w_{\infty}$  and non-linear $W_{\infty}$ algebras).
\end{description}

\subsection{Integrable model associated with $LWA_{+\infty}$ algebra}
\label{lkp}

 Define the affine vertex of $\widehat{sl_N}$ as $\dst
 a_n^0=\prod_{i=1}^{N-1}{{(a_n^i)}^{-1}}$.
The corresponding screening operator associated with the imaginary root of
$sl_N$ is
$$
Q_0= \sum_{n\in {\bf Z}}a_n^0
$$
Differential $\hat Q=Q_0+Q=\sum_{j=0}^N Q_j$ may be considered as
the hamiltonian of $\widehat{sl_N}$ - Toda system.
According to definitions of the work \cite{FF2},
Space of Integrals of Motion of this system is defined as  an intersection
\be
\label{affkernels}
\dst Ker(ad_{Q_0})\cap Ker(ad_{Q_1})\cap\ldots\cap
Ker(ad_{Q_{N-1}})\cap \frac{\Pi_0}{\partial \Pi_0}
\ee
The word {\em integrals} is encoded in the last intersection because of
obvious isomorphism
$$
\frac{\Pi_0}{\partial \Pi_0}\,\cong\,\Pi^{int}_0\leftarrow\Pi_0:\sum_n
$$
Before describing the space \Ref{affkernels}, let us take look at a
simpler  problem. It is
almost a trivial statement, that a system associated with the
pair ob brackets $\{,\}_1$ and $\{,\}_2$ is integrable, with infinite
number of conservation laws in involution. One just have to have {\em two}
 integrals, commuting under both brackets. The simplest choice is \cite{BC1}
\bean
&&I^{(1)}= \sum_n W_n^{(2)}\\
&&I^{(2)}= \sum\left({\left(W_n^{(2)}\right)^2\over 2}+
W_n^{(2)}W_{n+1}^{(2)}-W_n^{(3)}\right)
\eean
The subsequent procedure is obvious: using the bi-hamiltonian
structure, one can easily obtain the whole series of conservation laws
in involution by the recursive procedure. The answer for any $N$
(essentially, including
$N=\infty$) can be found in \cite{BC2}. We rewrite it here
for completeness. For given $N$, the
series is given by
\be
\label{ints}
\dst
I^{(k)}_N\,=\,\frac{1}{k}{\rm Tr}\left( {\cal L}_N^k\right)
\ee
where Lax matrix $L_N$ is defined through its inverse
\be
\label{Lax}
\left({\cal
    L}_N\right)^{-1}_{n,m}=\delta_{n,m+1}-W_n^{(2)}\delta_{n,m}+W_n^{(3)}
   \delta_{n,m-1}-\ldots +(-1)^{N-1}W_n^{(N)}\delta_{n,m-N+1}
\ee
Introduce the translation matrix $\dst
\Lambda_{n,m}=\delta_{n,m-1}$ and diagonal matrices
$\dst W^{(i)}_{n,m} = W^{(i)}_n\delta_{n,m}$. In
the compact notations $L$-operator \Ref{Lax} has the form
\be
{\cal L}_N=\Lambda\cdot\frac{1}{1 - L\cdot\Lambda +
  W^{(3)}\cdot\Lambda^2 -\dots +(-1)^{N-1}W^{(N)}\Lambda^{N-1}}.
\ee
It really is the $L$-operator of our dynamical system, because the
evolution equations can be written in the form
\be
\label{lax-repr}
{\d{\cal L}\over \d t_p}= [\BA_p,\,{\cal L}],
\ee
where $\BA^{(p)}_N=({\cal L}_N^p)_+$.
Now let us return to our original problem of description of the space
\Ref{affkernels}. Clearly, all the integrals described above
commute with $\dst Q=\sum_{j=0}^N Q_j$ even {\em locally} by the
construction \Ref{kernels}. In addition, direct calculation shows,
that {\em they also commute with} $Q_0$. The last step to make is to notice,
that of $N$ vertex operators corresponding to  $\widehat{sl_N}$ we
needed  only $N-1$ corresponding to simple roots to construct $LWA_N$
generators and integrals of
motion. In principle, we could pick up {\em any} $N-1$ vertex operators, and
follow the same steps. Eventually, the space of integrals of motion
for $\widehat{sl_N}$ Toda system  can be described in terms of
generating functions as
\be
\label{generfun}
R_{\widehat{SL_N}}(\lambda_1,\ldots,\lambda_N)=\sum_{i=1}^NR_{SL_N}^{(i)}
(\lambda_i)
\ee
where $\dst R_{SL_N}^{(i)}(\lambda_i)=\sum_{s=0}^{\infty}I^{(s)}\lambda_i^s$
is the generating function for the conservation
laws of the lattice $N$-KdV hierarchy, associated with the roots
$\{\#1,\#2,\dots,\#i-1,\#i+1,\ldots,\#N\}$.

Finally,  it is quite obvious that all the formulae and constructions
above apply directly to the case of $N=\infty$.




$M_j=\sum_{l=1}^j m_l$, $M_0=0$.


\subsection{Embedding of the Lattice NLS into the Lattice KP hierarchy}

It can be proved that evolution of the previously defined \l fields
$$
M_n^p={e_n f_{n+p}\over h_n h_{n+1}\ldots h_{n+p-1}}
$$
is consistent with lattice KP hierarchy \Ref{lax-repr} under the
following identification.
$$
L_{n,\,n+p}=(-1)^p M_n^{p+1}.
$$
Recall, however, that as defined by eq. \Ref{M-fields}, variables
$\{M_n^p\}_{p=1}^\infty$ are not functionally independent.
There is a set of quadratic relations, like  $M_n^2 M_{n+1}^2=M_n^3
M_{n+1}^1$, which may be interpreted as Plucker relations of some
Grassmanian. Two independent generators of the whole family are
$M_n^0$ and $M_n^1$, which form FTV algebra.  Thus, one may understand the
lattice NLS embedding into the lattice KP as a two {\em
  non-abelian} field realization of lattice KP. To compare this with
the {\em abelian} two-field relaization, mentioned in the end of
Section \ref{lwinf}, we notice that FTV algebra \Ref{ftv} and lattice
$u(1)$ algebra \Ref{ftv-1}, when treated just as Poisson structures,
can be viewed as respectively {\em second} and {\em first} brackets
for the Volterra hierarchy \cite{FT-86, VV}.



\section{Concluding remarks}
\label{concl}

In this paper we have studied the \l analogs of various conformal theories
as well as their integrable perturbations. We have found that when
described in proper invariant terms, many of the well-known continuous
constructions have their match on the lattice. Besides, reiterating
the ideas from the Introduction, we have discovered, that on the
lattice analogues of conformal theories and integrable models admit
description in universal terms. We explicitly described for the first
time lattice analogues of the Drinfel'd-Sokolov reduction and of the Sugawara
construction. In the framework of the \l WZW the \l Sugawara
energy-momentum tensor was constructed. Then we described \l MB
system as a ``chiral perturbation'' of the \l WZW model by the field
of spin one. Evolution equations under the integrals of motion of this system
form the integrable \l NLS hierarchy. Finally, we found an embedding
of the \l NLS hierarchy to the \l KP hierarchy, which seems to be very
natural in
continuous case.

There is still an open
question  how to  give
the geometrical description of the \l MB system using the Lie group
cosets, in analogy with continuous case \cite{ABF}.

We described the Spaces of IM's for several integrable systems
discussed in the paper, using Lax representation and bi-hamiltonian
structure. It would be extremely interesting to compare the results
of  cohomological \cite{FE} and St.-Petersburg
approaches \cite{FV,F1} with our answers. Recently S. Kryukov calculated
first three integrals in the quasiclassical limit of \l sine-Gordon
theory \cite{K}, using the generating function from the paper
\cite{FV}.  After careful comparison, we
found that his integrals of motion can be expressed in terms of
certain linear combinations of our ones.

\medskip\noindent
{\bf Acknowledgments}

We are grateful to B.~Feigin, A.~Belavin and members of his Seminar at
 Landau  Institute:
A. Kadeishvili, S. Kryukov, M. Lashkevich, S. Parkhomenko, V. Postnikov and
Ya. Pugay.  A.B. thanks V.~Rubtsov, T.~Takebe, V.~Drinfeld,
S.~Novikov, I.~Krichever  and S.~Pakulyak for fruitful discussions and
valuable comments.
K. C. is indebted to  L. D. Faddeev and  L. Bonora for interesting
comments and C.~Teleman for cohomological help.


\end{document}